# Computational modeling of magnetoconvection: effects of discretization method, grid refinement and grid stretching

A. Yu. Gelfgat$^1$, O. Zikanov$^2$

$^1$*School of Mechanical Engineering, Faculty of Engineering, Tel-Aviv University, Ramat Aviv, Tel-Aviv 69978, Israel. e-mail: gelfgat@tau.ac.il*

$^2$*Department of Mechanical Engineering, University of Michigan - Dearborn, Dearborn, MI 48128, USA. e-mail: zikanov@umich.edu*

**Abstract**

The problem of natural convection in a laterally heated three-dimensional cubic cavity under the action of an externally imposed magnetic field is revisited. Flows at the Rayleigh number $Ra = 10^6$ and the Hartmann number $Ha = 100$, and three different orientations of the magnetic field are considered. The problem is solved using two independent numerical methods based on the second order finite-volume discretization schemes on structured Cartesian grids. Convergence toward grid-independent results is examined versus the grid refinement and near-wall grid stretching. Converged benchmark-quality results are obtained. It is shown that for convection flows with a strong magnetic field a steep, sometimes extremely steep, stretching near some of the boundaries is needed. Three-dimensional patterns and integral properties of the converged flow fields are reported and discussed. It is shown that the strongest magnetic suppression is yielded by the field directed along the imposed temperature gradient. The horizontal magnetic field perpendicular to the imposed temperature gradient stabilizes the main convection roll and leads to a flow with higher kinetic energy and heat transfer rate than in the non-magnetic case. Applicability of the quasi-two-dimensional model to natural convection flows in a box is discussed.

1. Introduction

The magnetoconvection, i.e. thermal convection in an electrically conducting fluid affected by a magnetic field, is found in many astro- and geophysical (e.g. in star atmospheres or liquid planetary cores) and technological (e.g. in liquid-metal components of future nuclear fusion reactors, semiconductor crystal growth, casting of steel or aluminum, liquid metal batteries) systems (see, e.g. [1-5]). The distinctive feature of such magnetohydrodynamic (MHD) flows is their modification by the Lorentz body force that appears as the result of the interaction between the magnetic field and the induced electric currents permeating the fluid. If the magnetic field is strong, the modification is quite profound. Its main elements are: *(i)* suppression of turbulence, *(ii)* anisotropy of the flow structures, which become elongated or even two-dimensional in the direction of the magnetic field, and *(iii)* development of thin MHD boundary layers near the walls (see [6] for a review).

It has been recently understood that the suppression of turbulence by the magnetic field does not necessarily mean that the flow acquires a simple laminar steady-state form. On the contrary, growth of the MHD-specific convection instability modes that have weak or zero variations along the magnetic field lines and, thus, are not suppressed, may lead to unsteady, essentially nonlinear and complex flow dynamics [7-16].

The MHD modification of the flow presents additional challenges to numerical modeling. The challenges become especially serious in the cases of strong imposed magnetic field identifiable as those with large values of the Hartmann number $Ha$, which we define in section 2 of the paper. The main source of the challenges is the numerical stiffness of the problem due to the strong separation between the smallest and largest length and time scales. As an example, the Hartmann and sidewall boundary layers developing near the walls, respectively, perpendicular and parallel to the magnetic field, have the typical thickness $\sim Ha^{-1}$ and $\sim Ha^{-1/2}$. In addition to carrying strong gradients of velocity and electric potential, the layers are the locations of the larger part of the electric currents flowing in the fluid.

The numerical simulation experience accumulated over the recent years has shown that fine resolution of the MHD boundary layers is essential for accuracy of a computational model. For the magnetoconvection flows, this has been convincingly demonstrated in [7,8,11]. Even moderately insufficient resolution in these areas, for example, by fewer than 6-7 grid points within the Hartmann layer, can lead to not just quantitative errors, but also to a qualitatively incorrect picture of the entire flow.

Another conclusion of the recent computational work is that simulations of high-$Ha$ MHD flows require special attention to the discretization schemes applied to the electromagnetic fields (electric potential and current) and Lorentz force field [17-22]. In the flows with electrically insulated walls, where the electric currents close entirely within the liquid, a conservative discretization leading to global conservation of electric charge and momentum appears to be essential. Lack of this conservation property may lead to unphysical features, such as spurious oscillations of velocity near the walls, and numerical instability [17]. The conservative discretization was realized in the framework of finite-difference and finite-volume methods on staggered [17,20] and collocated [18,19,22] grids. The resulting algorithms were successfully applied to flows with $Ha$ up to 1000 (see, e.g., [18-20,7,8,10,11,23]). The situation appears to be different in flows with electrically conducting walls, where electric charge is not globally conserved within the fluid. As shown in [21], the conservation property is less important and may even have a slight detrimental effect on the scheme's stability in such flows.

The modeling of magnetoconvection in the case of a strong magnetic field was identified in [24] as one of the key problems of computational MHD. The configuration of natural convection in a laterally heated rectangular box with a uniform horizontal magnetic field was proposed as a possible benchmark. The choice was justified by the characteristic magnetoconvection phenomena, such as near-wall vertical jets, expected in this configuration [12] as well as by relevance to the design of liquid metal blankets for nuclear fusion reactors [4].

This paper continues the work on development of computational methods for high-$Ha$ MHD flows. The model problem of convection in a cubic box with lateral heating and an imposed magnetic field of three different orientations is solved. This is an extended version of the benchmark problem suggested in [24]. Similar solutions were attempted, on much lower level of numerical fidelity, in [25,26]. We obtain high-accuracy three-dimensional solutions, which may serve for future benchmarking of MHD codes. The effects of grid refinement, grid clustering, and various approaches to discretization are explored. The three-dimensional flow patterns are discussed and visualized using the novel approach proposed in [27,28].

## 2. Formulation of the problem

### 2.1. Formulation of the problem

We consider a natural thermal convection flow in a three-dimensional cubic box shown in Fig. 1. The length of the cube side is $L$. The vertical walls at $x = 0, L$ are kept at constant values of temperature, $T_{hot}$ and $T_{cold}$, respectively, while those at $z = 0, L$ and $y = 0, L$ are perfectly thermally insulated. The fluid has electrical conductivity $\sigma$, kinematic viscosity $\nu$, thermal diffusivity $\alpha$, and density $\rho = \rho_0[1 - \beta(T - T_{cold})]$, where $\beta$ is the thermal expansion coefficient and $\rho_0$ is the fluid density at temperature $T_{cold}$. The system is affected by a constant and homogeneous externally generated magnetic field of magnitude $B_0$ directed along one of the coordinate axes (see Fig. 1). We assume that the magnetic Prandtl and Reynolds numbers are very small, so that the quasi-static approximations can be used, in which the flow-induced perturbations of the magnetic field are neglected in comparison with the imposed magnetic field in the expressions for the Ohm's law and Lorentz force [6]. Denoting the unit vector in the direction of the magnetic field by $\boldsymbol{b}$, the dimensionless governing equations in the Boussinesq approximation read

$$\frac{\partial \boldsymbol{v}}{\partial t} + (\boldsymbol{v} \cdot \nabla)\boldsymbol{v} = -\nabla p + Gr^{-1/2}\Delta \boldsymbol{v} + \theta \boldsymbol{e}_z + HaGr^{-1/2}\boldsymbol{J} \times \boldsymbol{b}, \tag{1}$$

$$\Delta \varphi = Ha \nabla \cdot (\boldsymbol{v} \times \boldsymbol{b}), \tag{2}$$

$$\boldsymbol{J} = -\nabla \varphi + Ha(\boldsymbol{v} \times \boldsymbol{b}), \tag{3}$$

$$\nabla \cdot \boldsymbol{v} = 0 \tag{4}$$

$$\frac{\partial \theta}{\partial t} + (\boldsymbol{v} \cdot \nabla)\theta = Pr^{-1}Gr^{-1/2}\Delta \theta. \tag{5}$$

where the dimensionless parameters of the problem are the Hartman number

$$Ha = B_0 H (\sigma/\rho_0 \nu)^{1/2}, \tag{6}$$

the Grashof number

$$Gr = g\beta(T_{hot} - T_{cold})H^3/\nu^2, \tag{7}$$

and the Prandtl number

$$Pr = \nu/\alpha. \tag{8}$$

In (1)-(6), the non-dimensional temperature is $\theta = (T - T_{cold})/\Delta T$, with the temperature scale defined as $\Delta T = T_{hot} - T_{cold}$. The length, velocity, time, pressure, electric potential, and electric current are rendered dimensionless using, respectively, the box size $L$, free-fall speed $U = (g\beta \Delta T L)^{1/2}$, $L/U$, $\rho_0 U^2$, $UB_0 L/Ha$, and $\sigma U B_0/Ha$.

The walls allow no slip:

$$\boldsymbol{v} = 0 \quad \text{at all the walls.} \tag{9}$$

The walls are perfectly electrically insulated. As already specified, they are perfectly thermally insulated with the exception of the two opposing vertical walls maintained at constant temperatures. The boundary conditions for the dimensionless temperature and electric potential are:

$$\theta = 1, \quad \frac{\partial \varphi}{\partial x} = 0 \quad \text{at } x = 0, \tag{10}$$

$$\theta = 0, \quad \frac{\partial \varphi}{\partial x} = 0 \quad \text{at } x = 1, \tag{11}$$

$$\frac{\partial \theta}{\partial y} = 0, \quad \frac{\partial \varphi}{\partial y} = 0 \quad \text{at } y = 0,1, \tag{12}$$

$$\frac{\partial \theta}{\partial z} = 0, \quad \frac{\partial \varphi}{\partial z} = 0 \quad \text{at } z = 0,1. \tag{13}$$

## 3. Computational approach

One of the goals of the study is to cross-verify the results using two independent computational methods for MHD flows. Brief accounts of the methods are provided in this section. Detailed descriptions are available in the references given below.

### 3.1 Method I

The method is based on the finite-volume discretization scheme described in [29,30]. The electromagnetic part of the scheme has been added for the present study. The method includes direct calculation of steady states done by Newton iterations as in [31] and integration of the governing equations in time as in [30].

The semi-implicit projection algorithm is based on the second-order backward differentiation formula for the time derivative [29,30]. This requires solution of six elliptic (Poisson and Helmholtz) equations for the three velocity components, pressure, temperature, and electric potential.

The spatial discretization is conducted on a structured, non-uniform, Cartesian grid in the manner illustrated in Fig. 2a. The fully staggered arrangement similar to that in [17] is utilized. The central points of the cells $(x_i, y_j, z_k)$ are used to approximate pressure and temperature. The velocity components are approximated at the centers of the faces: $u_{i+\frac{1}{2},j,k}$, $v_{i,j+\frac{1}{2},k}$, $w_{i,j,k+\frac{1}{2}}$. The central points of the cell edges are used to approximate the electric potential. This part of the arrangement is done differently for different directions of the magnetic field. As an example, Fig.

2a illustrates the case of $\boldsymbol{b} = \boldsymbol{e}_x$, in which we use $\varphi^x_{i,j+\frac{1}{2},k+\frac{1}{2}}$. Similarly, the cases of $\boldsymbol{b} = \boldsymbol{e}_y$ and $\boldsymbol{b} = \boldsymbol{e}_z$ require, respectively, $\varphi^y_{i+\frac{1}{2},j,k+\frac{1}{2}}$ and $\varphi^z_{i+\frac{1}{2},j+\frac{1}{2},k}$. Only one of the fields is computed and used as the potential $\varphi$ in each of the specific configurations of the magnetic field considered in this paper. A superposition of all three fields has to be used at an arbitrary $\boldsymbol{b}$ (this situation requiring interpolation of the Lorentz force components is not considered further in our discussion).

The discretization of the momentum, temperature and pressure equations follows the classical staggered grid scheme and, therefore, consistent and conservative (see, e.g., [32] ). For the electromagnetic part, the fully staggered arrangement allows us to build a consistent discretization on a compact stencil, which does not require interpolation and conserves the electric charge exactly.

We will illustrated the scheme for the case of $\boldsymbol{b} = \boldsymbol{e}_x$ shown in Fig. 2a. The $y$- and $z$-components of the electric current are evaluated as

$$J_y|_{i,j,k+\frac{1}{2}} = -\frac{\varphi^x_{i,j+\frac{1}{2},k+\frac{1}{2}} - \varphi^x_{i,j-\frac{1}{2},k+\frac{1}{2}}}{y_{j+\frac{1}{2}} - y_{j-\frac{1}{2}}} + w_{i,j,k+\frac{1}{2}}, \qquad (14)$$

$$J_z|_{i,j+\frac{1}{2},k} = -\frac{\varphi^x_{i,j+\frac{1}{2},k+\frac{1}{2}} - \varphi^x_{i,j+\frac{1}{2},k-\frac{1}{2}}}{z_{k+\frac{1}{2}} - z_{k-\frac{1}{2}}} - v_{i,j+\frac{1}{2},k}. \qquad (15)$$

The right-hand side of the potential equation (2) is computed as

$$Ha\nabla \cdot (\boldsymbol{v} \times \boldsymbol{b})|_{i,j+\frac{1}{2},k+\frac{1}{2}} = \frac{w_{i,j+1,k+\frac{1}{2}} - w_{i,j,k+\frac{1}{2}}}{y_{j+1} - y_j} + \frac{v_{i,j+1,k+1} - v_{i,j+1,k}}{z_{k+1} - z_k}. \qquad (16)$$

The combination of the discretized gradient components in (14) and (15) and divergence in (16) gives the consistent second-order approximation of the Laplacian in the left-hand side of (2) on a compact stencil. Finally, the components of the Lorentz force in the momentum equation are computed immediately at the respective centers of the cell faces as

$$(\boldsymbol{J} \times \boldsymbol{b})_y|_{i,j+\frac{1}{2},k} = J_z|_{i,j+\frac{1}{2},k}, \qquad (\boldsymbol{J} \times \boldsymbol{b})_z|_{i,j,k+\frac{1}{2}} = -J_y|_{i,j,k+\frac{1}{2}}. \qquad (17)$$

We note that at $\boldsymbol{b} = \boldsymbol{e}_x$ the $x$-component of the force is zero, and the current component $J_x$ does not appear in the problem.

To resolve the boundary layers, especially the thin Hartmann layers developing at the walls perpendicular to the magnetic field, the grid points need to be strongly clustered towards the boundaries. The clustering is achieved via the coordinate transformation

$$x = 0.5 + 0.5 \frac{tanh[s(\xi - 0.5)]}{tanh(0.5s)}, \qquad (18)$$

where $\xi$ is the transformed coordinate, in which the grid is uniform, and $s$ is the stretching parameter that determines the degree of clustering. Larger $s$ means smaller distances between neighboring grid points near the boundaries and larger distances in the central part of the box.

It is well known that steep grid stretching causes numerical difficulties, which are connected to the extremely small distances between neighboring grid nodes. In particular, the iterative methods of solution of elliptic equations may lose convergence. Our computational experience shows that with the large stretching parameters used below, approximately at $s \geq 5$, the most effective multigrid and Krylov-subspace based iterative solvers fail to converge. This difficulty is overcome by using the tensor-product-factorization (TPF) and tensor-product-Thomas (TPT) solvers, as described in [29]. These solvers yield analytical solution to within the computer arithmetic precision. Additionally, the required amount of arithmetic operations remains the same for all the possible values of the governing parameters, which makes the solvers especially attractive for calculations at large Reynolds or Grashof numbers (see [29] for numerical examples). However, for the steep stretching applied here, an additional effort is needed. Thus, the one-dimensional eigenvalue decompositions, needed for TPT and TPF, are calculated with quadruple precision, after which only their double-precision part is used. This allows us to have 16 correct digits in the computed eigenvalues and eigenvectors, which further reduces the numerical errors. Furthermore, the Thomas algorithm suffers from numerical instability in the case of Neumann boundary conditions [33], which is the case for pressure and electric potential equations. To avoid this problem, a more computationally demanding TPF algorithm is used for calculation of the pressure and the potential.

### 3.2 Method II

The method uses the discretization scheme first introduced as scheme B in [22]. The scheme was expanded to include the effects of heat transfer and thermal convection and implicit treatment of temperature diffusion and viscous terms in [7,8,10]. The method was validated in comparison with experimental data and thoroughly tested for accuracy and efficiency in simulations of MHD flows at high Reynolds, Grashof, and Hartmann numbers (see, e.g. [7,8,10,11,23,34]). The scheme is valid for an arbitrary three-dimensional non-uniform magnetic field (see [7] for an example of such computations).

The method solves the unsteady governing equations in the form nearly identical to (1)-(5). The only difference is that the equations (2) and (3) do not include the factor $Ha$ in the right-hand sides due the absence of the factor $Ha^{-1}$ in the typical scales for the electric potential and current. The steady states are found as results of convergence of time-dependent solutions. The time discretization is of the second order and based on the backward-difference scheme with explicit two-layer approximation of nonlinear and force terms. The heat conduction and viscous terms are integrated implicitly to avoid the diffusive stability limits on the time step. The standard projection (fractional step) algorithm is used to satisfy incompressibility.

A structured Cartesian grid is used. The grid is non-uniform, with points clustered toward two sets of parallel walls according to the coordinate transformation (18). The grid remains uniform in the third direction due to the limitations imposed by the elliptic solver discussed below.

The spatial discretization scheme is of the second order and based on the principles proposed in [35,18,19]. It uses the collocated grid arrangement illustrated in Fig. 2b. All the flow variables and all the governing equations are approximated at the central points of the cells $(x_i, y_j, z_k)$. The method also uses the normal fluxes of velocity and electric current at the central points of the cell faces: $F_x|_{i+\frac{1}{2},j,k}$, $G_x|_{i+\frac{1}{2},j,k}$, $F_y|_{i,j+\frac{1}{2},k}$, $G_y|_{i,j+\frac{1}{2},k}$, $F_z|_{i,j,k+\frac{1}{2}}$, $G_z|_{i,j,k+\frac{1}{2}}$. The velocity fluxes $F$ are determined at the velocity correction step using the classical Rhie and Chow interpolation (see, e.g., [22]). The current fluxes are evaluated as proposed in [18]:

$$G_x|_{i+\frac{1}{2},j,k} = -\frac{\varphi_{i+1,j,k} - \varphi_{i,j,k}}{x_{i+1} - x_i} + \widetilde{(v \times b)}_x|_{i+\frac{1}{2},j,k}, \quad (19)$$

with analogous formulas for $G_y|_{i,j+\frac{1}{2},k}$ and $G_z|_{i,j,k+\frac{1}{2}}$. The wave above the second term in the right-hand side of (19) indicates that the term is the result of the linear interpolation from the cell-center points to the face centers. In order to calculate the Lorentz force, the components of the electric current $J$ at the cell centers are obtained from the current fluxes by linear interpolation.

As explained in detail in [18,19,22], the evaluation of divergence operator via the velocity or current fluxes leads to consistent approximation of the elliptic equations for pressure and electric potential on a compact stencil. The entire scheme is highly conservative. In the non-diffusive limit, the solution of the discretized equations exactly conserves mass, momentum, electric charge and internal energy, while kinetic energy is conserved with a dissipative error of the 3$^{rd}$ order.

The elliptic equations for pressure, electric potential, temperature, and velocity are solved by the direct method based on the combination of the Cosine Fast Transform in the direction, in which the grid is uniform, and cyclic reduction for the discretized two-dimensional elliptic problems for the transform coefficients.

## 3.3 Visualization of three-dimensional velocity fields

For visualization of three-dimensional velocity fields we implement the method proposed for incompressible flows in [27] . Divergence-free projections of velocity are made on the three sets of coordinate planes: (x,y), (y,z), and (x,z). We compute three projections $v_1$, $v_2$, $v_3$ of the velocity field $v$ on the subspaces formed by the divergence-free velocity fields having only two non-zero components. Consider, for example, the subspace of the coordinate plane (x,z) formed by all vectors $a$, such that $a = [a_x, 0, a_z]$, $\nabla \cdot a = \partial a_x/\partial x + \partial a_z/\partial z = 0$, and the components $a_x$ and $a_z$ satisfy the no-slip boundary conditions, where necessary. Denote the projection of the velocity field on this subspace as $v_1$ and the similar projections on the subspaces defined in planes (y,z) and (x,y) as $v_2$ and $v_3$, respectively. Thus, we obtain three vector fields, each of them having only two non-zero components, and each component being a three-dimensional scalar function:

$$v_1 = \begin{bmatrix} u_1(x,y,z) \\ 0 \\ w_1(x,y,z) \end{bmatrix}, \quad v_2 = \begin{bmatrix} 0 \\ v_2(x,y,z) \\ w_2(x,y,z) \end{bmatrix}, \quad v_3 = \begin{bmatrix} u_3(x,y,z) \\ v_3(x,y,z) \\ 0 \end{bmatrix}. \tag{20}$$

The two-dimensional divergence of each vector field vanishes:

$$div(v_1) = div_{(x,z)}(v_1) = \frac{\partial u_1}{\partial x} + \frac{\partial w_1}{\partial z} = 0 \tag{21}$$

$$div(v_2) = div_{(y,z)}(v_2) = \frac{\partial v_2}{\partial y} + \frac{\partial w_2}{\partial z} = 0 \tag{22}$$

$$div(v_3) = div_{(x,y)}(v_3) = \frac{\partial u_3}{\partial x} + \frac{\partial v_3}{\partial y} = 0. \tag{23}$$

This allows us to define a vector potential for each of the vector fields, which has only one non-zero component:

$$v_1 = \nabla \times \Psi_1; \quad \Psi_1 = \left(0, \Psi_y(x,y,z), 0\right), \tag{24}$$

$$v_2 = \nabla \times \Psi_2; \quad \Psi_2 = (\Psi_x(x,y,z), 0, 0), \tag{25}$$

$$v_3 = \nabla \times \Psi_3; \quad \Psi_3 = \big(0, 0, \Psi_z(x, y, z)\big). \tag{26}$$

Evidently, the three-dimensional function $\Psi_y(x, y, z)$ coincides with the streamfunction of $v_1$ in each plane $y = const$ and can be interpreted as an extended streamfunction. Similar interpretations are valid for $\Psi_x$ and $\Psi_z$. As a result, the fields $v_1, v_2$ and $v_3$ are tangent to the corresponding vector potential isosurfaces, and can be interpreted as divergence-free projections of the velocity field on the coordinate planes. Arguments for uniqueness of these projections, and different methods to compute them are given in [27,28].

Examples of convective flows in a laterally heated cube visualized by this technique can be found in [27]. In these examples the isosurfaces of $\Psi_y$ correspond to the main convection roll, in which the fluid ascends along the hot wall and descends along the cold one. The additional three-dimensional motion represented as two pairs of antisymmetric circulations in the $(y, z)$ and $(x, y)$ planes and is depicted by the isosurfaces of potentials $\Psi_x$ and $\Psi_z$.

## 4. Results

The simulations are performed at the same non-dimensional parameters as in [26]: a cubic cavity, the Prandtl number $Pr = 0.054$, the Rayleigh number $Ra = GrPr = 10^6$, and the Hartmann number $Ha = 100$, except for the test case, in which the magnetic field is absent. The orientation of the magnetic field along the $x$, $y$, or $z$ axis is indicated in the following discussion by the use of, respectively, $Ha_x$, $Ha_y$, or $Ha_z$.

The computed flows are analyzed using three-dimensional distributions and one-dimensional profiles of flow variables, and the integral parameters: the total kinetic energy

$$E_{kin} = \frac{1}{2} \int_0^1 \int_0^1 \int_0^1 v^2 \, dx\, dy\, dz, \tag{27}$$

and the Nusselt number

$$Nu = -\int_0^1 \int_0^1 \left(\frac{\partial \theta}{\partial x}\right)_{x=0} dy\, dz. \tag{28}$$

All the solutions with the magnetic field are found to be laminar and time-independent. This has been determined using the Method I and Method II by computing time-dependent solutions for a sufficiently long time to assure convergence to asymptotic steady states. These states obtained in the way just described or via the Newton's iterations with Method I are visualized and analyzed later in this section.

We note that the absence of time-dependency is not surprising from the physical viewpoint at such a combination of $Ra$ and $Ha$. It is attributed to the suppression of velocity fluctuations by the magnetic field.

Test simulations have been performed at zero magnetic field. Here, the flow demonstrates irregular oscillations characterized by several dominant dimensionless frequencies varying between approximately 0.01 and 0.1. The flow averaged in time over the time interval of 100 units is shown in Fig. 3. The isotherms show that temperature boundary layers develop near the boundaries $x = 0,1$. The distribution of $\Psi_y$ in Fig. 3(b) shows the main convection roll with the axis in the $y$-direction. It can also be observed that the secondary structures in the $(y, z)$ and $(x, y)$ planes are shifted towards the boundaries $x = 0,1$ and $z = 0,1$, respectively. The time-averaged integral parameters are $E_{kin} = 0.0218$ and $Nu = 6.593$. Note that for these parameters the calculations of [26] on the noticeably coarser grid arrived to the flow with $Nu = 7.2$.

### 4.1 Integral flow characteristics

The integral flow characteristics for all computed flows are listed in Tables 1-3. The grid size and the degree of grid stretching are varied in wide ranges: $100^3$ to $250^3$ and $1 \leq s \leq 12$ for the Method I, and $64^3$ to $256^3$ and zero stretching (a uniform grid) to $s = 6$ for the Method II. The grid stretching is chosen to be the same in all three directions in the Method I. In the Method II, the grid is always uniform in one direction and stretches differently (always stronger along the magnetic field) in the other two directions.

Our first conclusion is that convergence to grid-independent results is achieved for all three orientations of the magnetic field and for both the computational methods. The asymptotic accurate values that can be determined with a reasonably high certainty are listed in Table 4. At the same time, the convergence behavior varies significantly with the choice of the computational method as well as with the orientation of the magnetic field.

Fast convergence is observed at $Ha_x = 100$ or $Ha_z = 100$, i.e., when the magnetic field is parallel to the imposed temperature drop or vertical (see Tables 1 and 3). Even the crudest grids ($100^3$ points for Method I and $64^3$ points for Method II) give results within few percent of the asymptotic values if sufficiently strong grid stretching is used. The grids with $150^3$ points and $s = 6$ (Method I) and $128^3$ points and $(s_x, s_z) = (4,3)$ or $(3,4)$ for, respectively, $Ha_x = 100$ or $Ha_z = 100$ (Method II) appear sufficient for accurate simulations.

The convergence is noticeably slower in the case of the spanwise magnetic field, i.e., at $Ha_y = 100$ (see Table 2). For the Method I, the calculations with the weakest grid stretching ($s = 1$) produce quite wrong results. The accuracy improves dramatically on grids with strong stretching ($s = 9$ and $s = 12$) and appears to reach an acceptable level on the grids of $150^3$ points at $s = 12$ or $200^3$ points at $s = 9$. The situation is better in the case of the Method II, where the results obtained on the uniform grids are not very far from the asymptotic values. The grid of $128^3$ points and $(s_x, s_y) = (3,4)$ appears sufficient. Nevertheless, the convergence is slower than at the other orientations of the magnetic field.

Our results can be compared with those of [26], which shows $Nu = 4.766$ for $Ha_x = 100$, $Nu = 7.135$ for $Ha_y = 100$, and $Nu = 6.002$ for $Ha_z = 100$. We attribute the substantial difference observed in all three cases to the crude grid ($64^3$ or $128 \times 40^2$ uniformly distributed points) and the non-conservative discretization used in [26].

The effect of the magnetic fields of various orientations on the kinetic energy and heat transfer is summarized in Table 4, where the results are compared with those at $Ha = 0$. Considering the fact of suppression of velocity fluctuations by an imposed magnetic field, one would expect reduction of $E_{kin}$ and $Nu$ in the MHD flows. This is observed in our results when the magnetic field is aligned with the temperature drop or oriented vertically. In comparison to the non-magnetic case, the kinetic energy is reduced approximately three-fold at $Ha_x = 100$ and two-fold at $Ha_z = 100$. Strong reduction of $Nu$ is also observed.

The effect of the magnetic field is clearly different when the field is in the spanwise direction, i.e., at $Ha_y = 100$. We see that both $E_{kin}$ and $Nu$ increase in comparison to the non-magnetic case. The reasons for such a seemingly counterintuitive behavior will become clear when we consider the transformation of the flow structure.

## 4.2 Scalar fields

The effects of the magnetic field and grid parameters on the distributions of temperature $\theta$ and electric potential $\varphi$ are discussed in this section.

For the temperature, a nearly horizontal shape of the isotherms in the interior of the box indicates strong convective mixing (see the non-magnetic case illustrated in Fig. 3 as an example). This is not observed in the flows at $Ha_z = 100$ (see Fig. 4) and $Ha_x = 100$ (not shown). The suppression of the convective mixing is also revealed by the reduced values of $Nu$ shown Table 4. The situation is different in the flow with $Ha_y = 100$, which demonstrates high degree of mixing resulting in nearly horizontal isotherms in the middle of the domain (see Fig. 5) and increased value of $Nu$.

The effect of the grid size and stretching on the temperature distribution is observed for all three orientations of the magnetic field. Fig. 4 illustrates the effect for the case of $Ha_z = 100$. For different grids and stretching parameters we observe different steepness of the isotherm θ=0.5 in the central part of the cavity, as well as different deformation of the isotherms θ=0.1 and 0.9 in the boundary regions. Again, we observe that for the Method I, steep stretching ($100^3$ grid, s=9) yields better results than a refined grid with an insufficient stretching ($250^3$ grid, s=1). This effect becomes even more pronounced in the case of the spanwise magnetic field, directed along the y-axis (see Fig. 5).

Similar grid-dependency can be seen in the patterns of isobars (not shown here). Even more profound effects of the grid parameters on the results are observed in the patterns of equipotential surfaces that are shown in Figs. 6-8 for the three magnetic field directions. In all three cases, the the structures of the potential field obtained on different grids, for different stretchings and by different numerical methods are qualitatively similar. Quantitatively, however, we see substantial deviations of the fields obtained by the Method I on weakly stretched grids with $s = 1$ from the converged results. This evidently leads to incorrectly computed electromagnetic force and, consequently, the flow and temperature fields. We see this sensitivity of the electric potential to the details of the numerical procedure as requiring special attention. The potential is routinely ignored in visualization and analysis of MHD solutions. As we have just demonstrated, the electric potential is a sensitive measure of solution accuracy and should be a part of benchmarking along with the temperature and velocity.

In the case of the spanwise magnetic field $Ha_y = 100$ the potential distribution is nearly two-dimensional (with weak variation along the magnetic field lines) with thin layers of sharp potential gradients near the walls at $x = 0,1$ and $z = 0,1$ (see Fig. 8). We also note that owing to the symmetry of the problem, the equipontential surfaces form patterns with rotational symmetry, so that in cases of the magnetic field along the $x$-, $y$-, and $z$-axis, the symmetry is with respect to rotation around the $x$-, $y$-, and $z$-centerlines of the box, respectively.

The results obtained by both the methods were also examined by comparing one-dimensional profiles of solution variables. Such profiles are poorly suited for visualization of the three-dimensional flow structure, but allow for a better demonstration of quantitative differences. An example is given in Fig. 9 for the temperature profiles. We clearly see that disagreement between the not converged and converged results is the smallest in the case of $Ha_x = 100$, when the magnetic suppression of the flow is the strongest. The disagreement is the largest in the case of the spanwise magnetic field $Ha_y = 100$, for which the suppression of the bulk flow is the weakest, however changes of the flow pattern are significant (see below). Here we would like to reiterate that, in agreement with the visualization in Fig. 4, the profile corresponding to the converged solution shows nearly uniform temperature distribution in the interior of the box and sharp temperature gradients near the walls at $x = 0,1$. This is consistent with the large value of $Nu$ reported for $Ha_y = 100$ in Table 4.

### 4.3. Velocity profiles

The calculated velocity fields are compared using one-dimensional profiles, which allows us to emphasize the computational problems related to the grid coarseness and insufficient stretching, as well as to illustrate the structure of the velocity boundary layers.

Profiles of the $x$-component of velocity are shown in Fig. 10. The main elements of the profiles are the two strong horizontal jets near the top and bottom walls $z = 0,1$ (see Fig. 10(a)-10(c)). The jets are a part of the main convection roll and observed for all the magnetic field directions, as well as for the non-magnetic convection flow. The profiles show antisymmetry with respect to the mid-plane of the box $z = 0$. Under the magnetic field effect, the boundary layers at $z = 0,1$ become thinner, as already noticed in the two-dimensional simulations [13]. As shown in Fig. 10(b), the vertical magnetic field leads to a nearly linear distribution of velocity between the jets. The spanwise magnetic field (see Fig. 10(c)) results in the flow, in which the horizontal velocity in the core of the box is practically zero, but the jets near the top and bottom walls are much stronger than in the other two cases.

The profiles in Figs. 10(d)-(f) reveal the complex structure of the jets by showing the distributions of the velocity component $u$ along the $y$-axis at $x = 0.5$, $z = 0.05$, i.e. approximately at the location of the strongest flow in the bottom jet. We see that at $Ha_x = 100$ and $Ha_z = 100$ the jets have complex three-dimensional shapes with higher velocity near the walls at $y = 0,1$. The amplitude of the variation of $u$ along $y$ is in both cases stronger than for the time-averaged profile obtained at $Ha = 0$ (not shown).

We do not observe noticeable convergence or stretching-dependency problems when the magnetic field is directed along the $x$-axis (Fig. 10(a) and 10(d)). These problems become more profound when the magnetic field is vertical (Fig. 10(b) and 10(e)), and really strong when the field is directed along the y-axis (Fig. 10(c) and 10(f)). Similarly to what was observed for temperature an electric potential, simulations based on the Method I and using grids with weak stretching, $s = 1$, do not yield the profiles of the velocity component $u$ close to the converged ones.

The structure of another part of the main convection roll, namely the vertical jets near the hot ($x = 0$) and cold ($x = 1$) vertical walls, is presented in Fig. 11. The jets are clearly visible in the profiles along the horizontal central line $y = z = 0.5$ (see Figs. 11(a)-(c)). The maximum velocity is about the same at $Ha_y = 100$ and $Ha_z = 100$, but about three times smaller at $Ha_x = 100$. The effect can be attributed not just to the general strong suppression of the flow at $Ha_x = 100$ documented earlier, but also to the widening of the jets and disappearance of the core with nearly zero vertical velocity (cf. Fig. 11(a) and Figs. 11(b),(c)). We also note that the boundary layers at $x = 0,1$ are narrowed by the magnetic fields of all three orientations in comparison to the non-magnetic case.

The structure of the vertical jet along the $y$-axis is illustrated in Figs. 11(d)-(f). As for the horizontal jets, we see high-amplitude variation revealing strong three-dimensionality of the flow structure at $Ha_x = 100$ and $Ha_z = 100$, and the nearly flat profile caused by the magnetic field at $Ha_y = 100$. Thin boundary layers are visible near the walls at $y = 0,1$ in three cases. As for the profiles in Fig. 10, comparison with the non-magnetic case reveals higher degree of three-dimensionality at $Ha_x = 100$ and $Ha_z = 100$ and thinner boundary layers.

The profiles in Fig. 11 generally confirm the conclusions made above in regard of the effect of grid size and stretching on the velocity distribution. The effect is relatively mild at $Ha_x = 100$ and $Ha_z = 100$, but becomes strong at $Ha_y = 100$ when the Method I is used (see Figs. 11(c),(f)).

The $y$-component of velocity $v$ is noticeably smaller than the other two. This component is not a part of the main convection roll, but the result of the flow's three-dimensionality, which necessarily develops owing to the pressure drop between the box center and the borders $y = 0,1$ as well as the Lorentz force. Its profiles (not plotted) show that this component is also characterized by the steep boundary layers developing near the $x, z = 0,1$ walls. The grid- and stretching dependencies of the $y$-velocity profiles are similar to those observed for the other velocity components.

### 4.4. Three-dimensional flow structure

The converged flows calculated at all the three orientations of the magnetic field are visualized in Figs. 12-14 in the same manner as for the flow at $Ha = 0$ in Fig. 3. These three-dimensional flow visualizations help us to better understand the effect of the magnetic field on the flow. For the purpose of possible future comparisons, the minimum and maximum values of the velocity potentials, as well as the plotted levels, are reported in the captions.

When the magnetic field is directed along the $x$-axis, i.e., parallel to the externally applied temperature gradient, we observe the already documented suppression of the convection flow. The amplitude of the variation of the velocity potentials decreases substantially in comparison with the non-magnetic case (cf. the data in the captions of Figs. 3 and 12). This is in agreement with the reduction of the Nusselt number and the total kinetic energy reported in Table 4. The vector potential patterns in Fig. 12 show thickening of the boundary layers at $x = 0,1$ (cf. Figs. 3c and 12c). The flow remains essentially three-dimensional, although the velocity gradients in the magnetic field direction are reduced in the core of the box. It can be also observed, that the vector potential patterns of Fig. 12 are partially similar to those reported in [27] for the flow at a larger Prandtl number. This recalls a certain similarity between the electromagnetic damping and increase of viscosity, also discussed in the previous studies.

The plots in Fig. 13 show that the vertical magnetic field also suppresses the flow, but not as strongly as the $x$-directed field. This is seen in the integral characteristics listed in Table 4, and by the difference between the maximum and minimum values of the velocity potentials listed in the captions of Figs. 3 and 13. Note that the main circulation rolls have qualitatively different shapes in the cases of the $x$- and $z$- directed fields (cf. Fig. 12(a) and Fig. 13(a)). We also observe that the secondary vortices near the walls at $x = 0,1$ are shifted toward these walls (see Fig. 13(c)), and the corresponding motion in the central part of the cube weakens. Contrarily, the secondary vortices forming near the walls at $z = 0,1$ are shifted towards the central part of the box, out of the boundary layers (see Fig. 13(d)).

In the case of the spanwise magnetic field $Ha_y = 100$, we observe the main circulation roll of nearly two-dimensional form with very weak gradients of velocity along the *y*-axis outside the Hartmann boundary layers (see Figs. 14(a)-(b)). The circulations in the $(y, z)$ and $(x, y)$ planes are shifted into the boundary layers adjacent to the boundaries $x = 0,1$ and $z = 0,1$, respectively. We note that weak 3D motion remains near these boundaries.

Fig. 14 illustrates the following explanation of the increase of the flow's kinetic energy and Nusselt number in the flow with $Ha_y = 100$ in comparison with the non-magnetic case (see Table 4). Unlike the fields oriented along the *x*- and *z*-axes, the spanwise magnetic field is aligned with the axis of the main convection roll and, thus, does not suppress this roll except by the viscous friction in the Hartmann boundary layers at *y*=0,1. At the same time, the magnetic field suppresses the secondary convection structures due to the substantial *y*-derivative of velocity associated with them. This stabilizes the main roll and reduces the kinetic energy transfer from it to smaller three-dimensional structures. As proven by our simulations, the total kinetic energy and heat transfer rate increase in the result of this transformation.

We note that strong coherent convection rolls aligned with an imposed magnetic field are observed in other convection flows at high $Ha$ [7,8,11,36]. In many cases, it leads to apparently paradoxical behavior characterized by high-amplitude temperature oscillations or, as in our case, increase of the kinetic energy and Nusselt number.

### 4.5. Applicability of the Q2D model to natural convection flows

As the last step of the analysis, we utilize the computed flow fields to consider the question of validity of the quasi-two-dimensional (Q2D) model of MHD flows [37] in the case of natural convection in a box. The reasoning and conclusions presented below are not entirely new, but the supporting contribution of comprehensive three-dimensional modeling is new and, we believe, valuable.

The Q2D model was originally derived for an isothermal flow along a duct with electrically insulated walls and imposed transverse magnetic field, but applied to many other configurations since then (see, e.g. [9,16,38] for convection flow examples). The formal conditions of the model's validity is that the Hartmann number $Ha$ and the Stuart number $N \equiv B_0^2 \sigma L / \rho_0 U = Ha^2/Re$, where $Re$ is the Reynolds number, are both much larger to one. In essence, the model assumes that under the action of a strong magnetic field, the flow acquires a state with velocity components virtually uniform along the field lines except for the exponential distributions within the thin Hartmann boundary layers. Wall-to-wall integration in the field direction produces equations for two-dimensional integrated variables with the electromagnetic effect reduced to linear friction at the Hartmann walls.

The model can only be rigorously derived if the walls of the flow domain are perfectly electrically insulated. The fact that it becomes invalid in convection flow with electrically conducting walls was highlighted in [15].

In our simulations, $Ha = 100$, while the value of $N$ can be estimated using $Re = Gr^{1/2}$ (following from the definition of the typical velocity and length scales in section 2) as $N$=2.32. This is not high enough for the flow to be in true Q2D state in the sense of [37], but sufficient to observe whether or not the transformation toward such a state occurs. The analysis is based on the computed three-dimensional flow structures and profiles, such as shown in Figs. 4-14.

The Q2D model is invalid when the magnetic field is in the $x$- or $z$-direction. In both the cases, the field is perpendicular the main convection roll, so it crosses either the vertical or horizontal pair of jets of opposite directions forming near the walls. Transformation of the flow into a state with velocity uniform along the magnetic field would require complete destruction of the jets and drastic change of the flow's topology. Our results as well as the results of the studies [39,40] conducted for the case of vertical magnetic field do not indicate that such a transformation is likely. Rather, further increase of the strength of the magnetic field is expected to narrow the jets and shift them toward the walls.

The applicability of the Q2D model is geometrically possible in the case of the magnetic field along the $y$-axis, i.e. along the axis of the main convection roll. As we have already discussed in section 4.4, the magnetic field stabilizes the roll and suppresses its three-dimensional secondary instabilities. Figs. 4, 8, 10(f), 11(f), and 14 show development of the core of the flow with weak gradients along the magnetic field lines. At the same time, we have already observed that weak three-dimensional motion remains active near the boundaries at $x = 0,1$ and $z = 0,1$ (see Figs 14(c),(d)). One can say that the pattern of the potential $\Psi_x$ (Fig. 14(c)) supports the validity of the Q2D approximation, while pattern of $\Psi_z$ (Fig. 14(d)) shows that there exists 3D motion near the horizontal boundaries. We conclude that the Q2D is likely to be applicable to the configuration with the spanwise magnetic field, although further studies based on the comparison between the Q2D and 3D results at higher $Ha$ and $N$ are necessary before the final conclusion can be made.

## 5. Concluding remarks

Numerical simulations of natural convection in a box with imposed magnetic fields of three different orientations were performed for the case of large Rayleigh and Hartmann numbers. Two numerical methods and grids of various sizes and degrees of near-wall clustering were used. The convergence to grid-independent results was achieved for both methods, but the details of convergence were found to vary significantly depending on the method and the orientation of the magnetic field. In particular, we have found that, while all the configurations require strong grid clustering toward the walls, the requirements are particularly stringent in the case of the spanwise (oriented horizontally along the axis of the main convection roll) magnetic field.

The two numerical methods demonstrated convergence to the same grid-independent solutions. At the same time, the convergence was achieved on smaller grids and with weaker wall clustering when the Method II based on the conservative discretization on a collocated grid was used. The effect was particularly pronounced in the case of the spanwise magnetic field. We do not have a convincing explanation of the difference between the performances of the two methods, but would like to note that similar differences were indicated by the tests conducted in [17] and [18], where the discretizations similar to those of our Methods I and II were used. The effect warrants further analysis, which can be considered as a part of the general study of performance of MHD codes for high-$Ha$ flows [24].

The analysis of the flow structure and integral properties has shown that the magnetic field parallel to the imposed temperature drop yields the strongest suppression of the flow and convection heat transfer. The suppression is weaker in the case of the vertical magnetic field. The spanwise magnetic field stabilizes the main convection roll and leads to increase of the flow's kinetic energy and Nusselt number in comparison to the non-magnetic case.

It is interesting to relate the transformation of the flow structure and the performance of the two numerical models. The fastest convergence and weakest dependence on the grid stretching is observed at $Ha_x = 100$. In the cases of vertical and spanwise magnetic fields, the flow develops thinner boundary layers, which makes the convergence slower and increases the demand for the stretching steepness. Observing the patterns of the temperature, electric potential and velocity fields we conclude that the steep stretching near the boundaries orthogonal to the magnetic field is crucial. Regarding the stretching near the other boundaries, we observe that a smoother stretching or even a uniform grid can yield faster converging results. Apparently, the optimal stretching is problem-dependent and should be fit for each problem separately.

Our final comment is on the applicability of the Q2D model to the natural convection flows in a box. The model is not applicable when the magnetic field is in the $x$- or $z$-direction, i.e. when it is perpendicular to the axis of the main convection roll. The flow approaches two-dimensionality in the case of the spanwise magnetic field. Accurate Q2D results are expected in this configuration at higher Hartmann and Stuart numbers.

Acknowledgments: Financial support was provided by the US NSF (Grant CBET 1435269) and the University of Michigan - Dearborn.

# Figure captions

**Figure 1.** Sketch of the considered flow configuration. Laterally heated three-dimensional cubic box under the effect of external magnetic field of three possible directions.

**Figure 2.** Illustrations of the grid arrangement systems used in the study. *(a)*, The fully staggered grid applied in the Method I in the case of the magnetic field $\boldsymbol{b} = \boldsymbol{e}_x$. *(b)*, The collocated grid applied in the Method II for an arbitrary magnetic field $\boldsymbol{b}$.

**Figure 3.** Three-dimensional temperature and velocity fields for the time-averaged flow at $Ra = GrPr = 10^6$ without magnetic field. Calculations are performed using the Method I on the grid of $150^3$ points with the stretching parameter $s = 3.7$. The isotherms $\theta = 0.1, 0.5,$ and $0.9$ are plotted in (a). The velocity is visualized in (b)-(d) by vector potentials of divergence-free velocity projections and projections of velocity vectors on selected planes. Maximum and minimum values of the velocity potentials are: (b) -0.0766, 0.000395; (c) -0.0209, 0.0193; (d) -0.0160; 0.0184. Levels plotted are: (b) -0.03; (c) ±0.001; (d) ±0.009.

**Figure 4.** Isotherms $\theta = 0.1, 0.5,$ and $0.9$ for $Pr$=0.054, $Ra$=$10^6$, $Ha_z$=100 calculated on different grids with different stretching by the Methods I and II.

**Figure 5.** Isotherms $\theta = 0.1, 0.5,$ and $0.9$ for $Pr$=0.054, $Ra$=$10^6$, $Ha_y$=100 calculated on different grids with different stretching by the Methods I and II.

**Figure 6.** Equipotential surfaces for $Pr$=0.054, $Ra$=$10^6$, $Ha_x$=100 calculated on different grids with different stretching by the Methods I and II. The levels $\varphi = 2.0, 4.0$ and $6.0$ are shown.

**Figure 7.** Equipotential surfaces for $Pr$=0.054, $Ra$=$10^6$, $Ha_z$=100 calculated on different grids with different stretching by the Methods I and II. The levels $\varphi = 2.0, 4.0$ and $6.0$ are shown.

**Figure 8.** Equipotential surfaces for $Pr$=0.054, $Ra$=$10^6$, $Ha_y$=100 calculated on different grids with different stretching by the Methods I and II. The levels $\varphi = 2.0, 4.0$ and $6.0$ are shown.

**Figure 9.** Temperature profiles along the line $y = z = 0.5$ in flows with three different orientations of the magnetic field, on different grids with different stretching computed by the Methods I and II.

**Figure 10.** Profiles of the *x*-component of velocity $u$ in flows with three different orientations of the magnetic field computed on different grids with different stretching by the Methods I and II. The profiles are drawn along the lines (a) – (c) $x = y = 0.5$, (d) – (f) $x = 0.05, z = 0.5$.

**Figure 11.** Profiles of the *z*-component of velocity $w$ in flows with three different orientations of the magnetic field computed on different grids with different stretching by the Methods I and II. The profiles are drawn along the lines (a) – (c) $x = y = 0.2$, (d) – (f) $x = 0.05, z = 0.5$.

**Figure 12.** Visualization of the velocity field for $Ha_x = 100$ by vector potentials of divergence free velocity projections. Maximum and minimum values of the velocity potentials are: (a,b) -0.0465, 0.000300; (c) ±0.00825; (d) ±0.00370. Levels plotted are: (a) -0.021; (b) -0.029; (c) ±0.003; (d) ±0.0022.

**Figure 13.** Visualization of the velocity field for $Ha_z = 100$ by vector potentials of divergence free velocity projections. Maximum and minimum values of the velocity potentials are: (a,b) -0.0584, 0.00504; (c) ±0.0251; (d) ±0.0105. Levels plotted are: (a) -0.0122; (b) -0.0266; (c) ±0.025; (d) ±0.007.

**Figure 14.** Visualization of the velocity field for $Ha_y = 100$ by vector potentials of divergence free velocity projections. Maximum and minimum values of the velocity potentials are: (a,b) -0.0510, 0.00571; (c) ±0.00642; (d) ±0.0101. Levels plotted are: (a) -0.04; (b) -0.048; (c) ±0.002; (d) ±0.006.

**Table 1.** The total kinetic energy $E_{kin}$ and Nusselt number $Nu$ calculated by the two methods on grids of various sizes and various degrees of stretching. The magnetic field is parallel to the *x*-axis, i.e., parallel to the applied temperature drop. $Ha_x = 100$.

*Method I*

| s | | 100 | 150 | 200 | 250 |
|---|---|---|---|---|---|
| 1 | $E_{kin}$ | 0.00498 | 0.00531 | 0.00553 | 0.00568 |
|   | $Nu$ | 4.236 | 4.300 | 4.342 | 4.372 |
| 3 | $E_{kin}$ | 0.00574 | 0.00595 | 0.00607 | 0.00614 |
|   | $Nu$ | 4.384 | 4.426 | 4.448 | 4.462 |
| 6 | $E_{kin}$ | 0.00637 | 0.006407 | 0.00642 | 0.00643 |
|   | $Nu$ | 4.503 | 4.511 | 4.514 | 4.516 |
| 9 | $E_{kin}$ | 0.00644 | 0.00645 | 0.00646 | 0.00646 |
|   | $Nu$ | 4.514 | 4.519 | 4.520 | 4.521 |
| 12 | $E_{kin}$ | 0.00644 | 0.00645 | 0.00646 | 0.00646 |
|    | $Nu$ | 4.508 | 4.517 | 4.519 | 4.521 |

*Method II*

| s | | 64 | 128 | 256 |
|---|---|---|---|---|
| Uniform | $E_{kin}$ | 0.00609 | 0.00636 | 0.00644 |
|         | $Nu$ | 4.503 | 4.514 | 4.530 |
| $s_x = 4$, $s_z = 3$ | $E_{kin}$ | 0.00641 | 0.00645 | 0.00646 |
|                      | $Nu$ | 4.530 | 4.525 | 4.524 |
| $s_x = 6$, $s_z = 4$ | $E_{kin}$ | 0.00642 | 0.00645 | 0.00646 |
|                      | $Nu$ | 4.537 | 4.526 | 4.521 |

**Table 2.** The total kinetic energy $E_{kin}$ and Nusselt number $Nu$ calculated by the two methods on grids of various sizes and various degrees of stretching. The magnetic field is parallel to the z-axis, i.e., parallel to the gravity force. $Ha_z = 100$.

*Method I*

| s | | 100 | 150 | 200 | 250 |
|---|---|---|---|---|---|
| 1 | $E_{kin}$ | 0.00895 | 0.00938 | 0.00966 | 0.00985 |
|   | $Nu$ | 5.181 | 5.227 | 5.260 | 5.283 |
| 3 | $E_{kin}$ | 0.00994 | 0.0102 | 0.0104 | 0.0105 |
|   | $Nu$ | 5.293 | 5.326 | 5.344 | 5.355 |
| 6 | $E_{kin}$ | 0.0108 | 0.0108 | 0.0109 | 0.0109 |
|   | $Nu$ | 5.387 | 5.393 | 5.396 | 5.398 |
| 9 | $E_{kin}$ | 0.0109 | 0.0109 | 0.0109 | 0.0109 |
|   | $Nu$ | 5.394 | 5.399 | 5.401 | 5.402 |
| 12 | $E_{kin}$ | 0.0108 | 0.0109 | 0.0109 | 0.0109 |
|    | $Nu$ | 5.388 | 5.397 | 5.400 | 5.401 |

*Method II*

| s | | 64 | 128 | 256 |
|---|---|---|---|---|
| Uniform | $E_{kin}$ | 0.0106 | 0.0109 | 0.0109 |
|         | $Nu$ | 5.523 | 5.430 | 5.408 |
| $s_x = 3$, $s_z = 4$ | $E_{kin}$ | 0.0109 | 0.0109 | 0.0109 |
|                      | $Nu$ | 5.423 | 5.408 | 5.404 |
| $s_x = 4$, $s_z = 6$ | $E_{kin}$ | 0.0109 | 0.0109 | 0.0109 |
|                      | $Nu$ | 5.427 | 5.409 | 5.403 |

**Table 3.** The total kinetic energy $E_{kin}$ and Nusselt number $Nu$ calculated by the two methods on grids of various sizes and various degrees of stretching. The magnetic field is parallel to the y-axis, i.e., horizontal and orthogonal to the applied temperature drop. $Ha_y = 100$.

*Method I*

| s | | 100 | 150 | 200 | 250 |
|---|---|---|---|---|---|
| 1 | $E_{kin}$ | 0.00807 | 0.00960 | 0.0112 | 0.0125 |
|   | $Nu$ | 7.070 | 4.989 | 5.000 | 5.543 |
| 3 | $E_{kin}$ | 0.0132 | 0.0158 | 0.0174 | 0.0186 |
|   | $Nu$ | 5.647 | 5.976 | 6.153 | 6.264 |
| 6 | $E_{kin}$ | 0.0228 | 0.0235 | 0.0239 | 0.0241 |
|   | $Nu$ | 6.600 | 6.644 | 6.665 | 6.678 |
| 9 | $E_{kin}$ | 0.0245 | 0.0247 | 0.0248 | 0.0248 |
|   | $Nu$ | 6.706 | 6.716 | 6.719 | 6.721 |
| 12 | $E_{kin}$ | 0.0244 | 0.0247 | 0.0248 | 0.0249 |
|   | $Nu$ | 6.700 | 6.718 | 6.722 | 6.723 |

*Method II*

| s | | 64 | 128 | 256 |
|---|---|---|---|---|
| Uniform | $E_{kin}$ | 0.0244 | 0.0245 | 0.0248 |
|   | $Nu$ | 6.901 | 6.787 | 6.731 |
| $s_x = 3$, $s_y = 4$ | $E_{kin}$ | 0.0255 | 0.0250 | 0.0250 |
|   | $Nu$ | 6.839 | 6.761 | 6.736 |
| $s_x = 4$, $s_y = 5$ | $E_{kin}$ | 0.0257 | 0.0251 | 0.0250 |
|   | $Nu$ | 6.849 | 6.763 | 6.748 |

**Table 4.** The estimated grid-independent values of the total kinetic energy $E_{kin}$ and Nusselt number $Nu$ for various orientations of the magnetic field. The time-averaged values at zero magnetic field are included for comparison.

| Configuration | $Ha = 0$ | $Ha_x = 100$ | $Ha_y = 100$ | $Ha_z = 100$ |
|---|---|---|---|---|
| $E_{kin}$ | 0.022 | 0.0065 | 0.025 | 0.011 |
| $Nu$ | 6.59 | 4.52 | 6.75 | 5.40 |

Figure

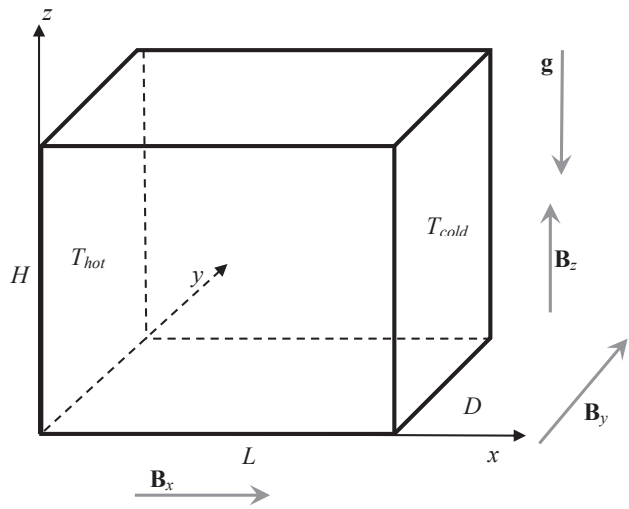

**Figure 1.** Sketch of the considered flow configuration. Laterally heated three-dimensional cubic box under the effect of external magnetic field of three possible directions.

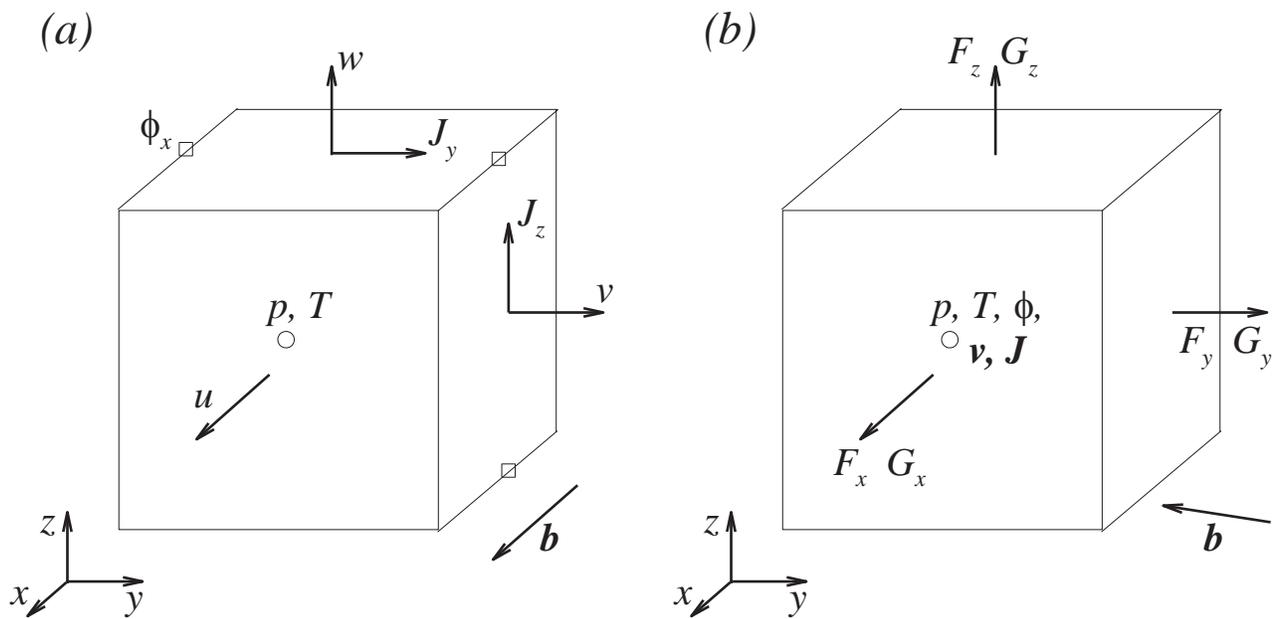

**Figure 2.** Illustrations of the grid arrangement systems used in the study. *(a)*, The fully staggered grid applied in the Method I in the case of the magnetic field $\boldsymbol{b} = \boldsymbol{e}_x$. *(b)*, The collocated grid applied in the Method II for an arbitrary magnetic field $\boldsymbol{b}$.

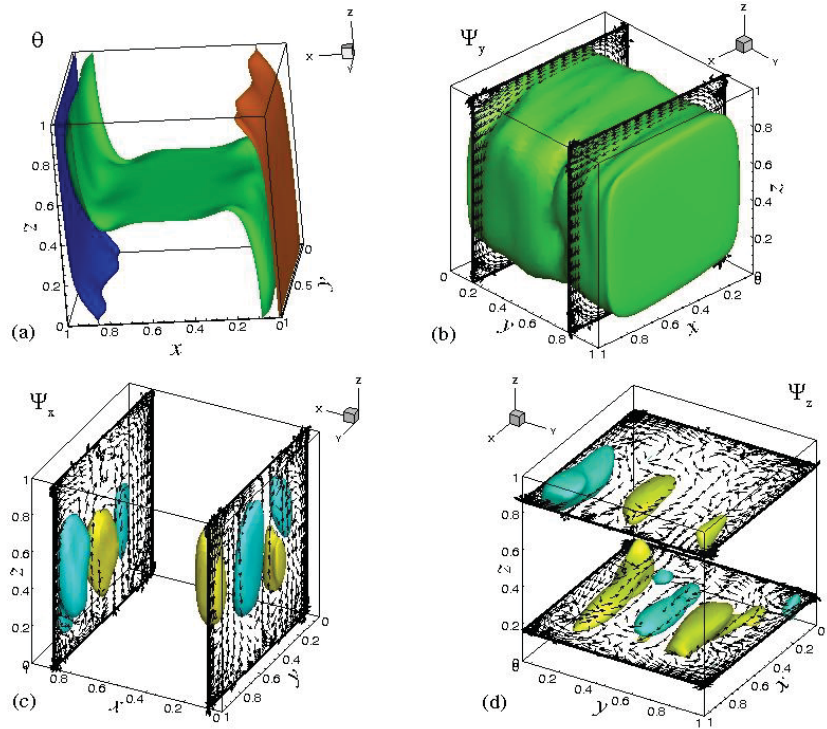

**Figure 3.** Three-dimensional temperature and velocity fields for the time-averaged flow at $Ra = GrPr = 10^6$ without magnetic field. Calculations are performed using the Method I on the grid of $150^3$ points with the stretching parameter $s = 3.7$. The isotherms $\theta = 0.1, 0.5$, and $0.9$ are plotted in (a). The velocity is visualized in (b)-(d) by vector potentials of divergence-free velocity projections and projections of velocity vectors on selected planes. Maximum and minimum values of the velocity potentials are: (b) -0.0766, 0.000395; (c) -0.0209, 0.0193; (d) -0.0160; 0.0184. Levels plotted are: (b) -0.03; (c) ±0.001; (d) ±0.009.

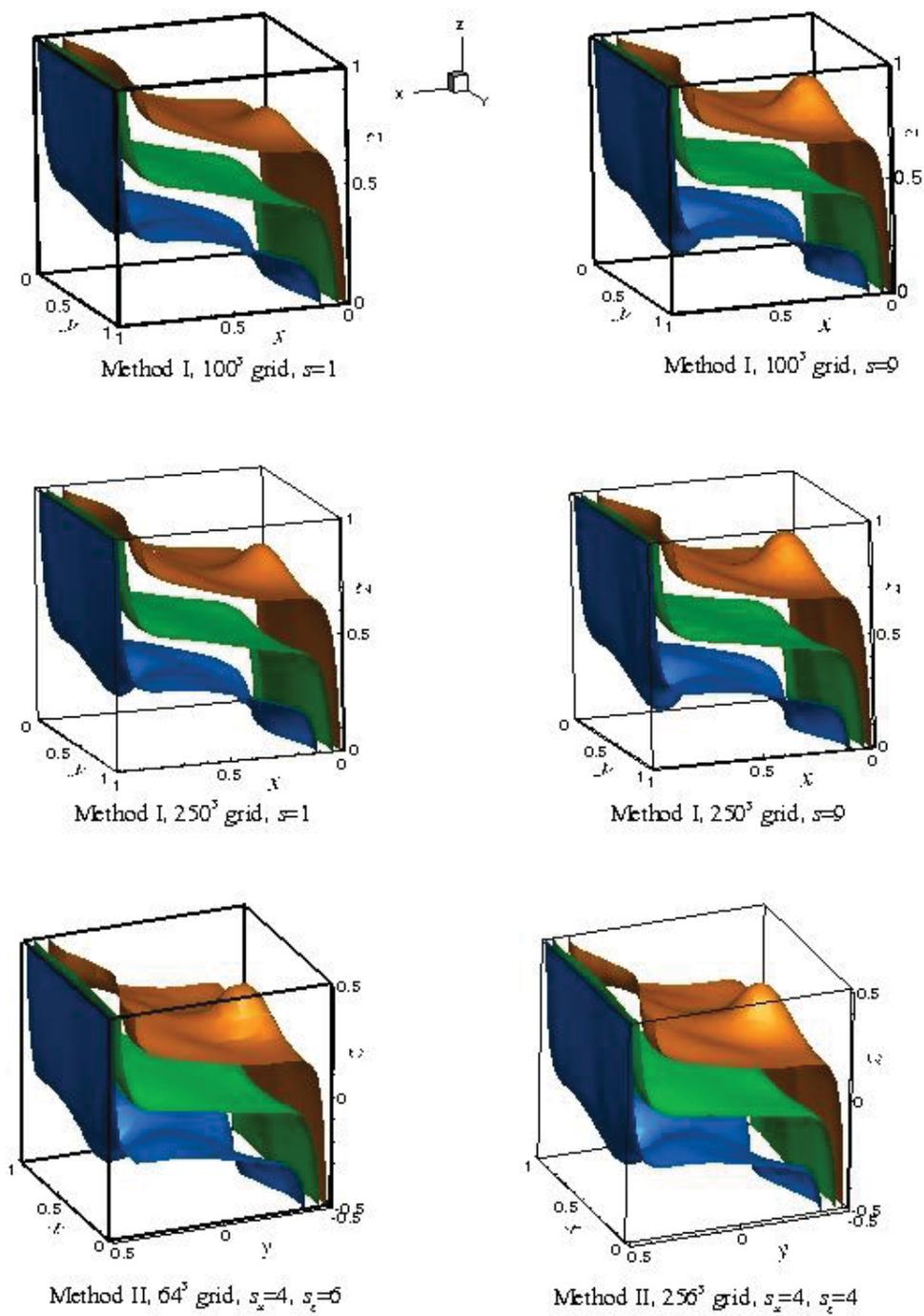

**Figure 4.** Isotherms $\theta = 0.1, 0.5,$ and $0.9$ for $Pr=0.054$, $Ra=10^6$, $Ha_z=100$ calculated on different grids with different stretching by the Methods I and II.

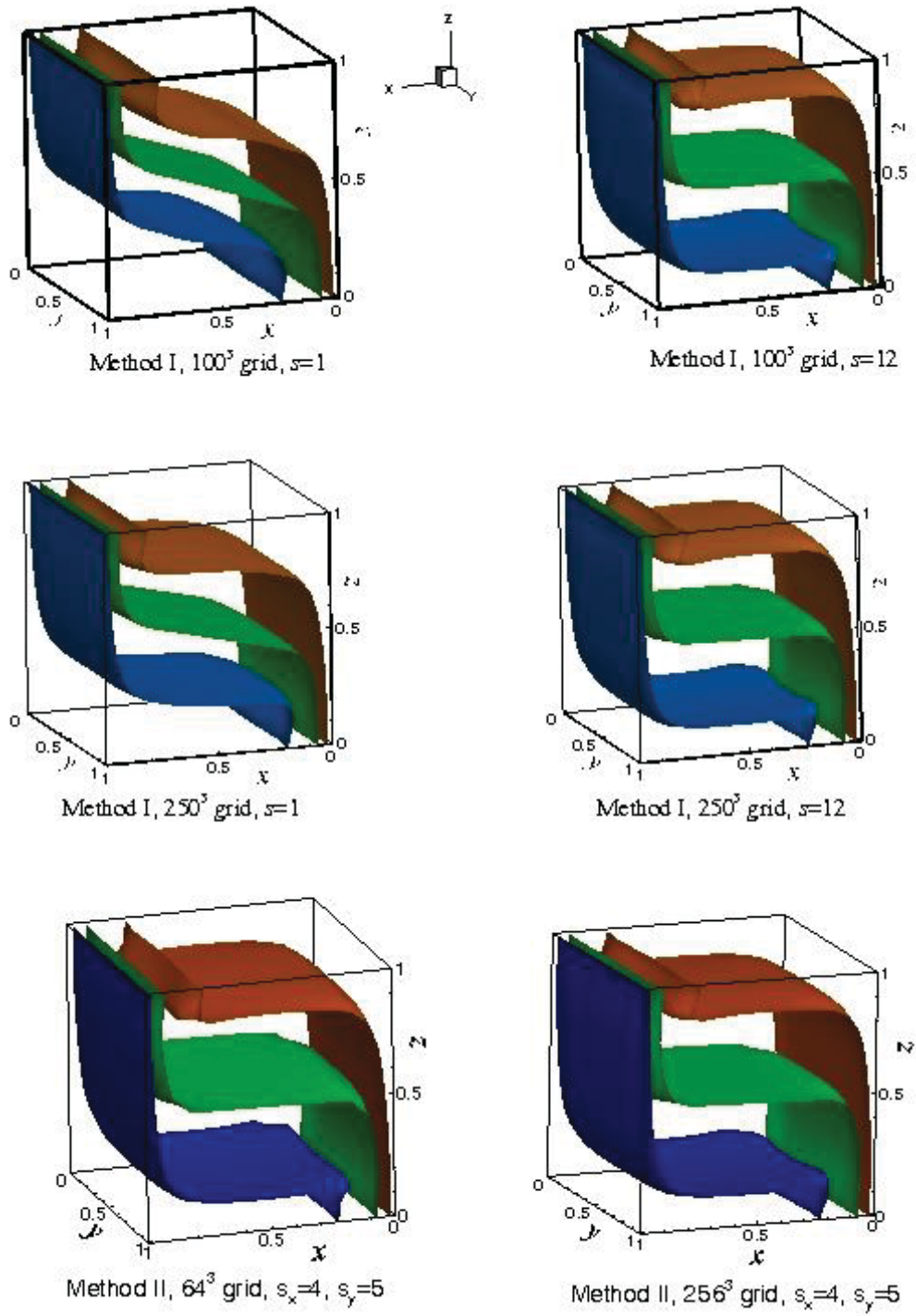

**Figure 5.** Isotherms $\theta = 0.1, 0.5,$ and $0.9$ for $Pr$=0.054, $Ra$=$10^6$, $Ha_y$=100 calculated on different grids with different stretching by the Methods I and II.

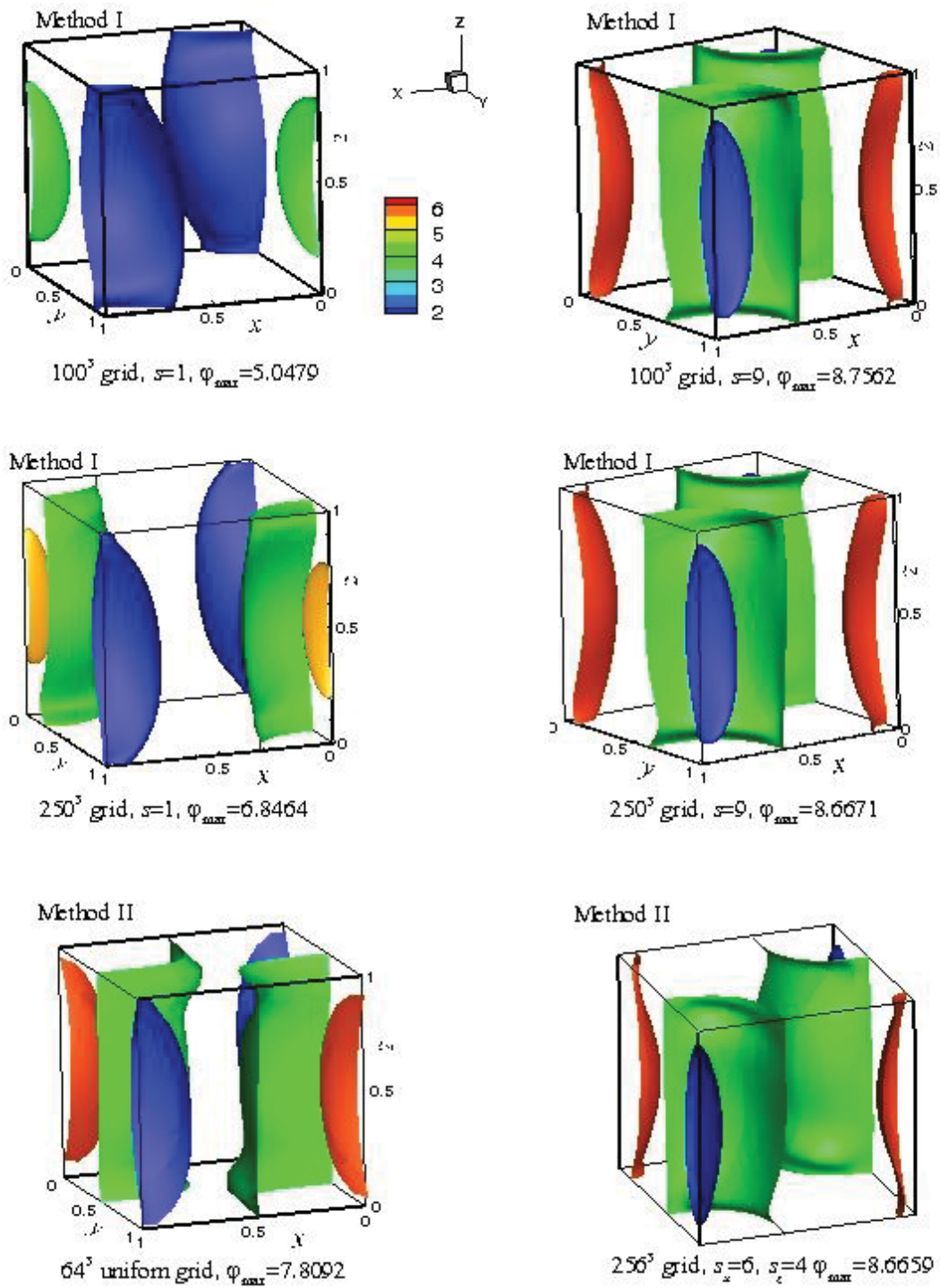

**Figure 6.** Equipotential surfaces for $Pr=0.054$, $Ra=10^6$, $Ha_x=100$ calculated on different grids with different stretching by the Methods I and II. The levels $\varphi = 2.0$, $4.0$ and $6.0$ are shown. In this figure and in figures 7 and 8, the free additive constant in the definition of the potential is fixed by setting the minimum value of $\varphi$ to zero.

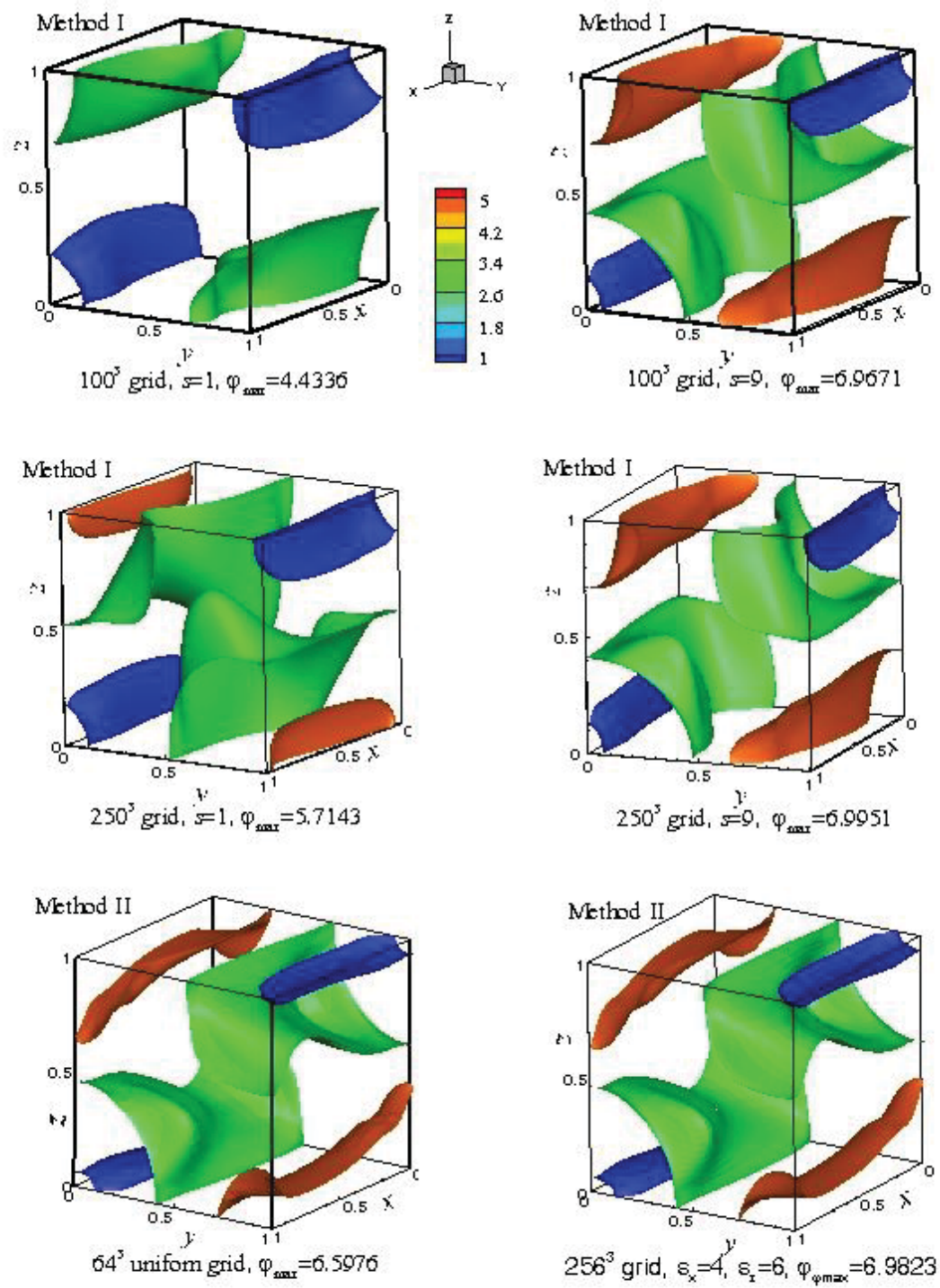

**Figure 7.** Equipotential surfaces for $Pr=0.054$, $Ra=10^6$, $Ha_z=100$ calculated on different grids with different stretching by the Methods I and II. The levels $\varphi = 2.0$, 4.0 and 6.0 are shown.

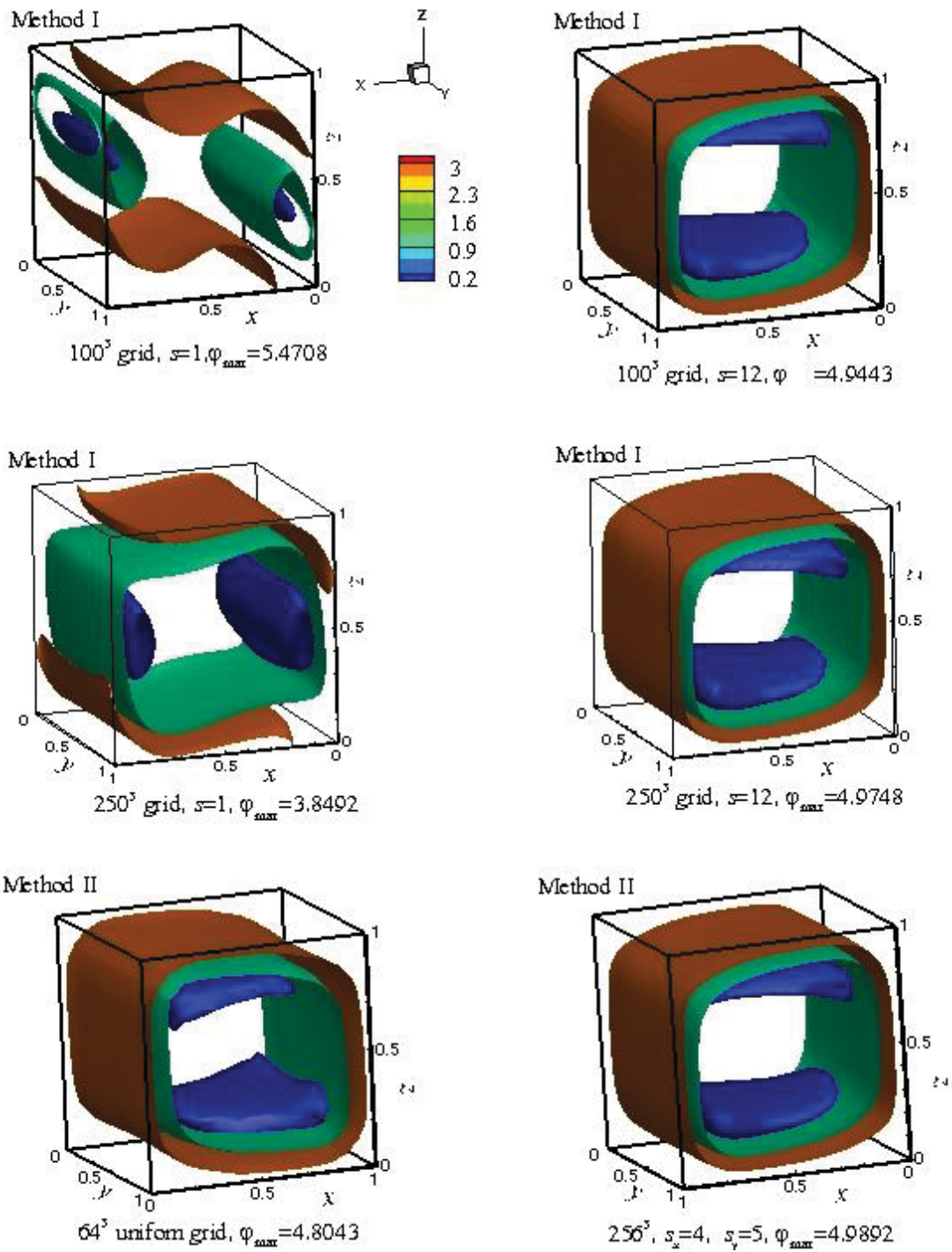

**Figure 8.** Equipotential surfaces for $Pr=0.054$, $Ra=10^6$, $Ha_y=100$ calculated on different grids with different stretching by the Methods I and II. The levels $\varphi = 2.0$, 4.0 and 6.0 are shown.

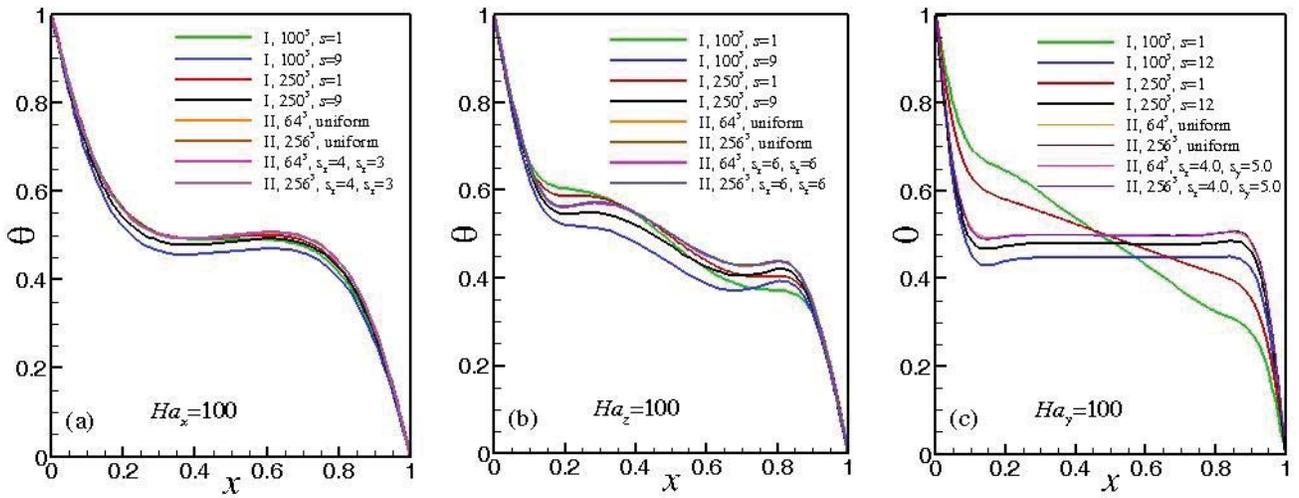

**Figure 9.** Temperature profiles along the line $y = z = 0.5$ in flows with three different orientations of the magnetic field, on different grids with different stretching computed by the Methods I and II.

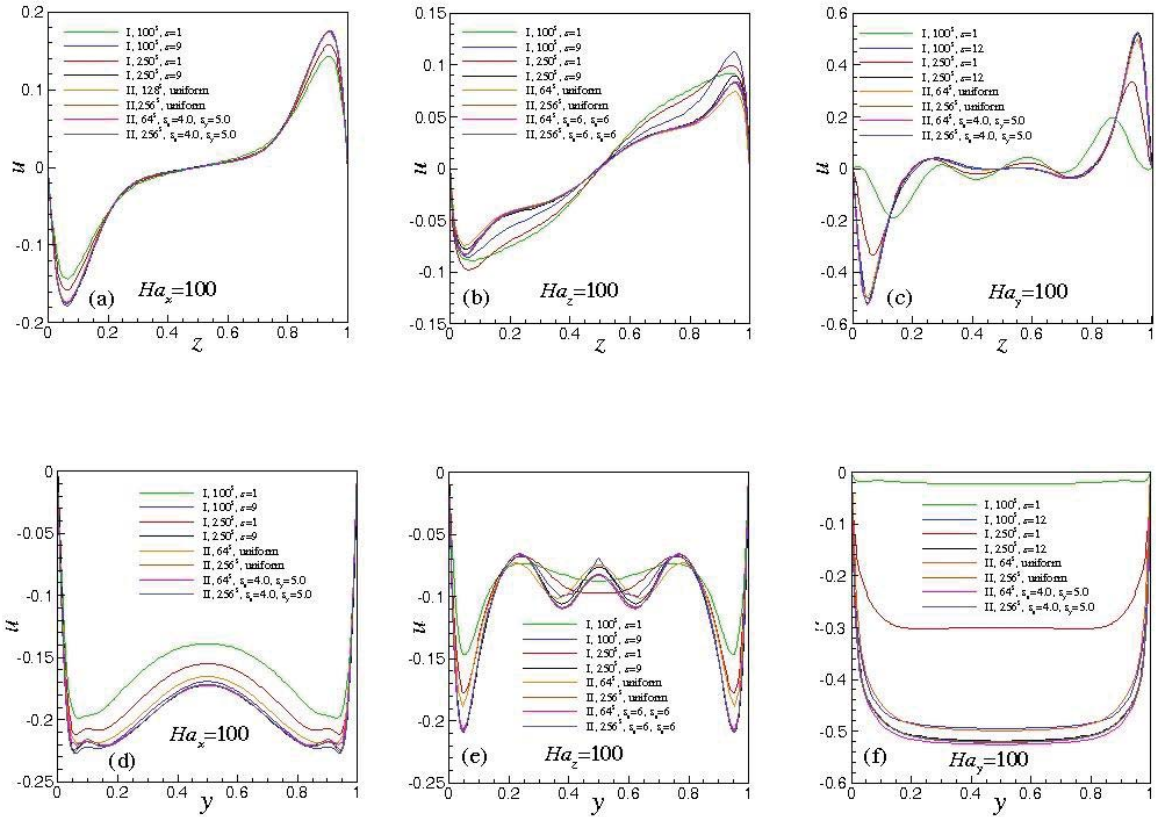

**Figure 10.** Profiles of the *x*-component of velocity $u$ in flows with three different orientations of the magnetic field computed on different grids with different stretching by the Methods I and II. The profiles are drawn along the lines (a) – (c) $x = y = 0.5$, (d) – (f) $x = 0.5, z = 0.05$.

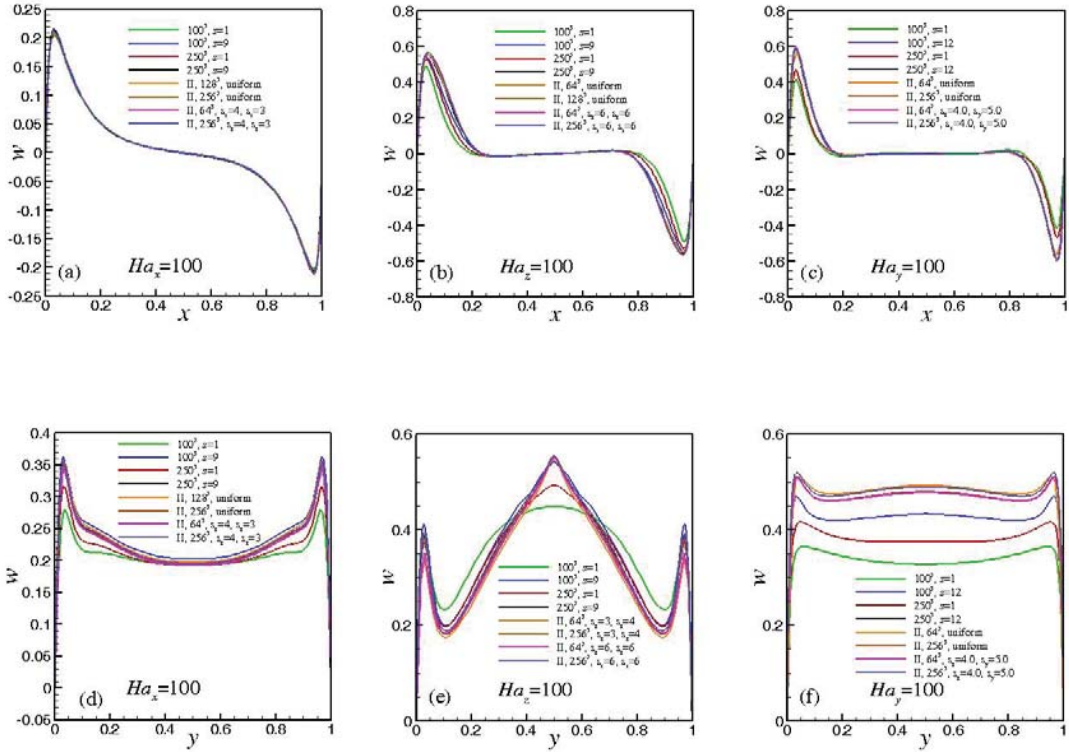

**Figure 11.** Profiles of the $z$-component of velocity $w$ in flows with three different orientations of the magnetic field computed on different grids with different stretching by the Methods I and II. The profiles are drawn along the lines (a) – (c) $y = z = 0.5$, (d) – (f) $x = 0.05, z = 0.5$.

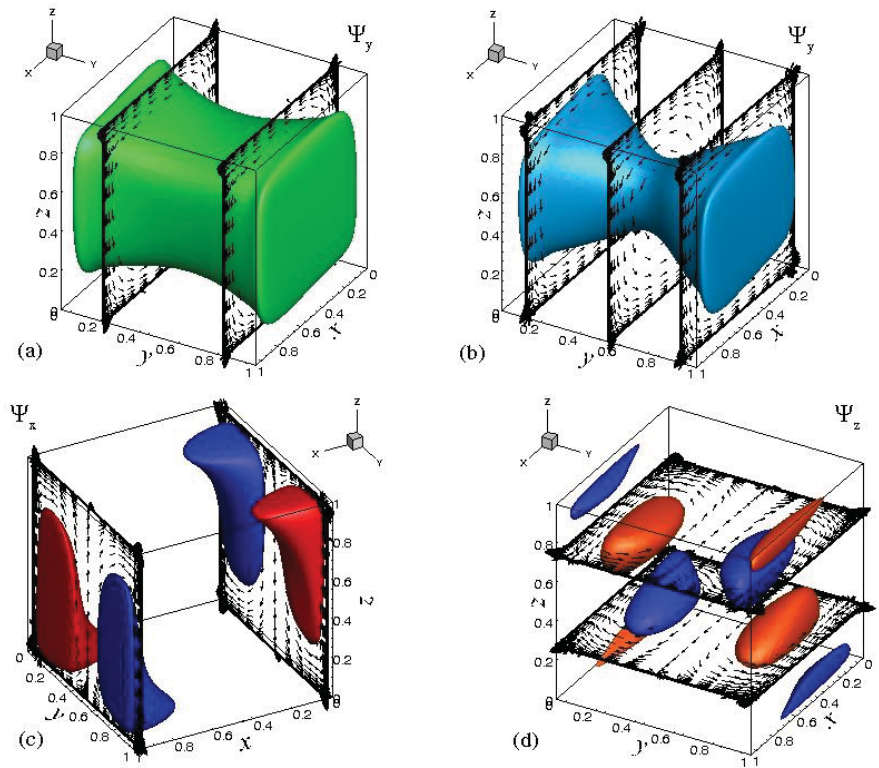

**Figure 12.** Visualization of the velocity field for $Ha_x = 100$ by vector potentials of divergence free velocity projections. Maximum and minimum values of the velocity potentials are: (a,b) -0.0465, 0.000300; (c) ±0.00825; (d) ±0.00370. Levels plotted are: (a) -0.021; (b) -0.029; (c) ±0.003; (d) ±0.0022.

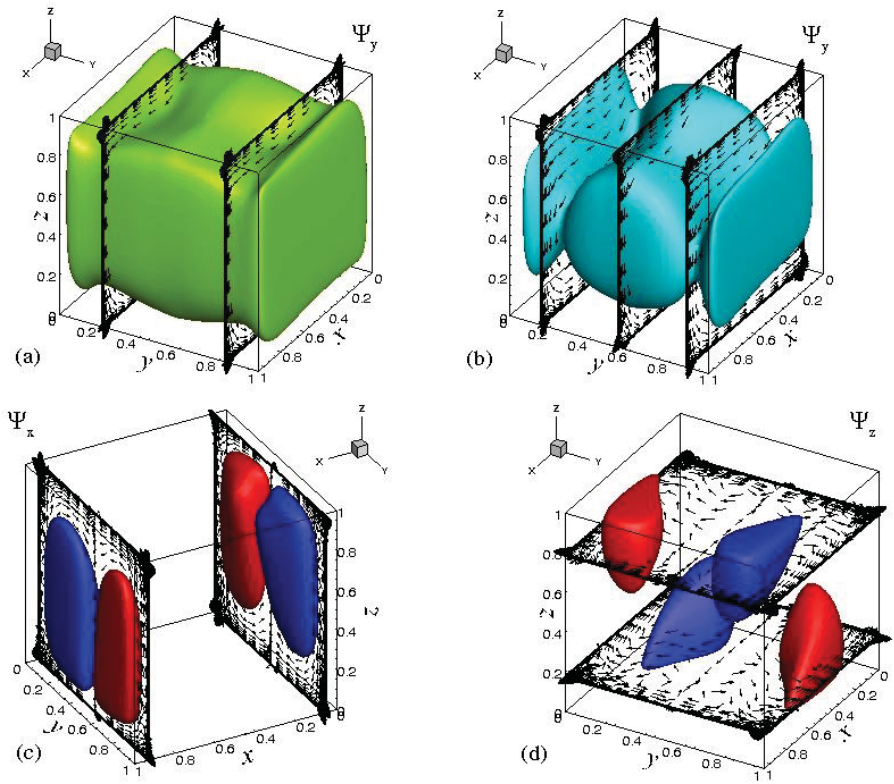

**Figure 13.** Visualization of the velocity field for $Ha_z = 100$ by vector potentials of divergence free velocity projections. Maximum and minimum values of the velocity potentials are: (a,b) -0.0584, 0.00504; (c) ±0.0251; (d) ±0.0105. Levels plotted are: (a) -0.0122; (b) -0.0266; (c) ±0.025; (d) ±0.007.

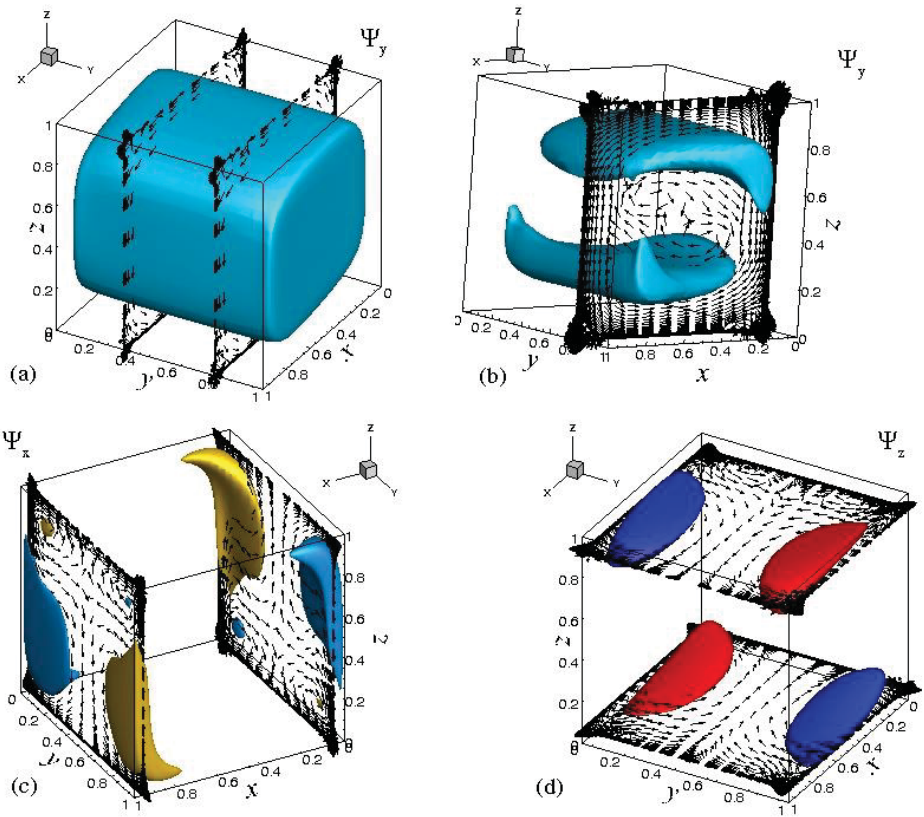

**Figure 14.** Visualization of the velocity field for $Ha_y = 100$ by vector potentials of divergence free velocity projections. Maximum and minimum values of the velocity potentials are: (a,b) -0.0510, 0.00571; (c) ±0.00642; (d) ±0.0101. Levels plotted are: (a) -0.04; (b) -0.048; (c) ±0.002; (d) ±0.006.